\documentclass[12pt]{article}

\usepackage[latin2]{inputenc}
\usepackage{amsmath}
\usepackage{amssymb}
\usepackage{amsfonts}
\usepackage{stmaryrd}
\usepackage{theorem}
\usepackage{graphicx}
\usepackage{epstopdf}
\usepackage[parfill]{parskip}    % Activate to begin paragraphs with an empty line rather than an indent
\usepackage{hyperref}
\hypersetup{pdftex,colorlinks=true,allcolors=blue}
\usepackage{hypcap}
\usepackage{algorithm}
\usepackage{algpseudocode} % algorithmicx package
\usepackage{authblk}
\usepackage{tabto}
\usepackage{wrapfig} % wrap text around figures
\usepackage{siunitx} % SI units.
\usepackage[titletoc,title]{appendix}
\usepackage{color}
\usepackage[bordercolor=white,backgroundcolor=gray!30,linecolor=black,colorinlistoftodos]{todonotes} % Highlighter 1.
\usepackage[framemethod=default]{mdframed} % Highlighter 2.

\newcommand{\keywords}[1]{\par\noindent{\small{\em Keywords\/}:\ \ #1}}
\newcommand{\mscclass}[1]{\par\noindent{\small{\em MSC class\/}:\ \ #1}}
\definecolor{myyellow}{HTML}{FFFFB3}
\mdfdefinestyle{mystyle}{hidealllines=true,backgroundcolor=myyellow,innerleftmargin=0pt,innerrightmargin=0pt,innertopmargin=0pt,innerbottommargin=0pt,skipabove=0pt,skipbelow=0pt}
\newcommand*{\me}{\mathrm{e}}

\newcommand*{\vr}{\ensuremath{\varrho}}

\newcommand*{\vt}{\ensuremath{\vartheta}}

\newcommand*{\vs}{\ensuremath{\sigma}}

\newcommand*{\vg}{\ensuremath{\gamma}}

\newcommand*{\N}{\ensuremath{\mathbb{N}}}

\newcommand*{\R}{\ensuremath{\mathbb{R}}}

\newcommand*{\mto}{\ensuremath{\rightarrow}}

\newcommand*{\md}{\mathrm{d}} % for integration dx etc.
\newcommand{\libeq}{\mathrel{\mathop:}=} % let it be equal
\newcommand*{\rar}{\ensuremath{\Rightarrow}}

\newcommand*{\kon}{\ensuremath{k_{\mathrm{on}}}}
\newcommand*{\koff}{\ensuremath{k_{\mathrm{off}}}}
\newcommand*{\meq}{\ensuremath{\mathrm{eq}}}
\newcommand*{\mdis}{\ensuremath{\mathrm{dis}}}

\newcommand*{\dcdt}{\ensuremath{\partial_t c}}

\newcommand*{\dt}{\ensuremath{\partial_t}}
\newcommand*{\dx}{\ensuremath{\partial_x}}
\newcommand*{\delt}{\ensuremath{\Delta t}}
\newcommand*{\delx}{\ensuremath{\Delta x}}
\newcommand*{\dely}{\ensuremath{\Delta y}}

\newcommand*{\dcdx}{\ensuremath{\partial_x c}}
\newcommand*{\ddcdx}{\ensuremath{\partial_x^2 c}}

\newcommand*{\Leq}{\ensuremath{\bar{L}}}
\newcommand*{\Teq}{\ensuremath{\bar{T}}}

\newcommand*{\Ceq}{\ensuremath{\bar{C}}}
\newcommand*{\xdet}{\ensuremath{x_{\mathrm{det}}}}
\newcommand*{\mIC}{\mathrm{IC}}
\newcommand*{\mBC}{\mathrm{BC}}

\newcommand*{\mmet}{\ \si{\metre}}
\newcommand*{\msec}{\ \si{\second}}
\newcommand*{\mmcmol}{\ \si{\cubic\metre/\mole}}
\newcommand*{\mmsVsec}{\ \si{\square\metre/\volt\second}}
\newcommand*{\mmssec}{\ \si{\square\metre/\second}}
\newcommand*{\mmolmc}{\ \si{\mole/\cubic\metre}}

\hypersetup{bookmarksnumbered}

\begin{document}

\title{A Fast Stable Discretization of the\\ Constant--Convection--Diffusion--Reaction Equations\\ of Kinetic Capillary Electrophoresis (KCE)}
\author{J\'{o}zsef Vass,\ \ Sergey N. Krylov\footnote{Corresponding author.}}
\affil{jvass@yorku.ca,\ \ skrylov@yorku.ca\\ \ \\ Centre for Research on Biomolecular Interactions\\ Department of Chemistry, York University\\ Toronto, ON, M3J 1P3, Canada}
\date{September 7, 2017}
\maketitle

\begin{abstract}
\noindent A discretization scheme is introduced for a set of convection--diffusion equations with a non-linear reaction term, where the convection velocity is constant for each reactant. This constancy allows a transformation to new spatial variables, which ensures the global stability of discretization. Convection--diffusion equations are notorious for their lack of stability, arising from the algebraic interaction of the convection and diffusion terms. Unexpectedly, our implemented numerical algorithm proves to be faster than computing exact solutions derived for a special case, while remaining reasonably accurate, as demonstrated in our runtime and error analysis.\\
%\ \\
\mscclass{65M08 (primary); 65M12, 35C05 (secondary).}
\keywords{Convection--diffusion equations, multimesh, stable discretization.}
\end{abstract}

\newpage
\tableofcontents

\newpage
\section{Introduction} \label{s01}

\subsection{Aims and Overview} \label{s0101}

The aim of this paper is to develop a numerical algorithm for solving a general system of equations -- the Constant--Convection--Diffusion--Reaction (CCDR) equations (Sec. \ref{s0201}) -- applicable to the experimental models of both Kinetic Capillary Electrophoresis (KCE; Sec. \ref{s0201}) and KSEC (Sec. \ref{s0504}). Both are reversible binding reactions in a long and narrow tube: a capillary or a separation column. The CCDR equations model chemical reactions between any number of substances, with an arbitrary reaction term, convected in a fluid and propagated by a constant electric field, describing the spacetime evolution of reactant concentrations. Though the convection is constant in our model, this does not alone eliminate the instability of standard discretization schemes which convection--diffusion equations are notorious for, but it does allow a change of variables which induces stability (Sec. \ref{s0301}).

The presented mathematical endeavour is fundamentally important for two directions of our ongoing scientific research in kinetic separation: (1) the accurate computer simulation of experiments under appropriate conditions; (2) the resolution of an experimental--computational inverse problem for determining kinetic rate constants -- both within the KCE framework. For these applications, a practical numerical solution method must be both stable and accurate, though for the inverse problem, a low runtime of this direct solver is also critical.

Previously, our simulations of KCE were generated with COMSOL, which employs a streamline upwind Petrov--Galerkin method to solve the general Nernst--Planck Equations (Secs. \ref{s0201} and \ref{s030101}). This method is neither globally stable however, as evidenced by its output, nor fast enough for resolving the aforementioned inverse problem, described below. Furthermore, the use of a readily available black box simulation software, like COMSOL, has obvious limitations, such as the lack of flexibility for modification or extension to a broader framework, like an inverse problem. By their very generality, such software are likely to be suboptimal for a specific purpose. Thus the introduced numerical algorithm is not only an accurate simulation tool, but also an efficient direct solver serving a larger vision within the KCE framework, as follows.

The KCE system of convection--diffusion--reaction equations includes kinetic rate constants in the reaction term, which can be viewed as parameters that the solution functions of this system of partial differential equations depend on. The efficient and accurate generation of such parametric solutions to the direct problem is necessary for the evaluation of the error target function at each iteration of some optimization algorithm, which minimizes this target function to resolve the inverse problem of determining the parameters that induce a given target signal \cite{vasskrylov2017kceinv}.

Finding the accuracy of a numerical solution is difficult in general, but for the KCE model, it can be compared with an exact solution for a simplified case, derived in Sec. \ref{s0302}. The error is analyzed in Sec. \ref{s0403}, and the runtimes of the direct solver in Sec. \ref{s0401} are compared in Sec. \ref{s0402}.

\subsection{The Physical Model} \label{s0201}

The equations of our physical model are deduced from the Nernst--Planck Equations, which express the conservation of mass of ions in a fluid medium under the influence of an electric field along a single spatial direction, while accounting for convection, diffusion, and reaction between the ions.

Define $c = (c_1,\ldots,c_N): \R_+^2\mto\R_+^N$ (where $\R_+\libeq [0,+\infty)$) as the spatiotemporal concentration of $N\in\N$ convected ions over $(t,x)$ points, $V:\R^2\mto\R^N$ as the velocities of their convection, $D:\R_+^2\mto\R_+^N$ as their diffusion, $R:\R_+^N\mto\R^N$ as a reaction term between the ions, $K\in\R^N$ as coefficients arising from various electromagnetic and thermodynamic constants \cite{site002}, and $E:\R^2\mto\R^N$ as the electric fields influencing the motion of the ions, while denote with $\cdot$ the Hadamard product of two vectors. To arrive at an intermediate form towards our equations, the Nernst--Planck equations \cite{probstein1994physicochemical, site002} reduce to the following vector PDE
\[ \dcdt + \dx(V\cdot c) = \dx(D\cdot\dx c) + R(c) + K\cdot\dx(E\cdot c) \]
with appropriate initial and boundary conditions that ensure the existence and uniqueness of a concentration vector solution.

For the experimental model of Kinetic Capillary Electrophoresis (KCE) \cite{berezovski2002nonequilibrium, petrov2005kinetic, okhonin2006plug, krylov2007kinetic}, the above equation reduces further due to certain features of the experimental setup. Specifically the convection velocities are constant $V\in\R^N$, the diffusion coefficients are constant $D\in\R_+^N$, and the electric field strength $E$ is constant as well. Thus a constant vector $v\libeq V - K\cdot E\in\R_+^N$ is introduced, and its components can be thought of as a case of constant convection in a convection--diffusion equation, suggesting the name Constant--Convection--Diffusion--Reaction (CCDR) equations. Therefore, they can be stated as
\[ \dcdt + v\cdot\dcdx = D\cdot\ddcdx + R(c) \]
again with with appropriate initial and boundary conditions \cite{saville1986theory, palusinski1986theory}.

In the formulation of the KCE model, there are three reactants and their concentration vector is denoted either as $c = (L, T, C):\R_+^2\mto\R_+^3$ representing the experimental ligand, target, and complex, or as $c = (A, B, C)$ in the simplified case of MASKE \cite{okhonin2010maske}. The velocities and diffusion coefficients are denoted as $v = (v_L, v_T, v_C)$ and $D = (D_L, D_T, D_C)$ respectively. In the general KCE model, the reaction term is
\[ R(c) = (-\kon LT + \koff C,\ -\kon LT + \koff C,\ \kon LT -\koff C): \R_+^2\mto\R^3. \]
Here $\kon,\ \koff\in\R_+$ are the rate constants of complex formation
and dissociation respectively. In MASKE, the concentration of the second ion $T$ is assumed to be constant, so the second equation is omitted and the reaction term becomes $R(c) = (-\kon LT + \koff C,\ \kon LT -\koff C)$. The point of this simplification is to enable us to find an exact solution \cite{okhonin2010maske}. Another simplification for the same reason, is to keep all three concentration components and simplify the reaction term only as $R(c) = (\koff C,\ \koff C,\ -\koff C)$, as in Sec. \ref{s0302} \cite{okhonin2004nonequilibrium}.

\subsection{Initial and Boundary Conditions} \label{s0202}

Since our goal is to give both a numerical and an exact solution to the above equations, the latter must be conveniently derivable from a Green function via function convolutions. So the initial and boundary conditions (IBC) of the CCDR equations are chosen to be idealizations of the initial concentration profiles measured in KCE experiments.

The initial conditions for the KCE equations specify the concentration profiles of the injected reactant plugs. A plug can be represented using various density functions $\vr:\R\mto\R_+,\ \int_{-\infty}^{+\infty} \vr = 1$. An identical substance amount is assumed within the plug -- regardless of the density function used -- to ensure that the amount of reacting molecules remains consistent for various injected concentration profiles. The total amount of molecules at concentration $c(x)$ in a capillary of constant radius $r$, at locations $x$ within cylinders of infinitesimal height $\md x$, is given by the integral
\[ \int_0^{\infty} c(x)\ \pi r^2\ \md x. \]
So for two concentration functions to give the same amount of molecules, the areas under them must be equal. To be consistent with our earlier work \cite{krylov2007kinetic}, this amount is set at $\pi r^2 l\bar{c}$, which arises from the special case of rectangular concentration profiles of heights $\bar{c}\in\R^N$ and plug length $l>0$. To rescale a density function $\vr$ as another density, the proper transformation is $(1/l)\vr(x/l)$, which retains a unit area under the curve. To ensure the standard amount of molecules, the area under the initial concentration profile $\mIC\libeq c(0,\cdot):\R_+\mto\R_+$ must equal $\bar{c}l$, implying the transformation $\bar{c}\ \vr(x/l)$.

This way our earlier table of initial and boundary conditions \cite{krylov2007kinetic} can be generalized to arbitrary plug densities as follows.
\vspace{-0.2cm}
\begin{center}
\bgroup
\def\arraystretch{1.25}
\begin{table}[H]
\caption{\label{s020201}Initial and boundary conditions for standard KCE methods.}
\vspace{0.3cm}\hspace{0.02cm}
\begin{tabular}{ | l | l | l | l | }
    \hline
    \textbf{Method} & \boldmath$c(0,x)$ & \boldmath$c(t,0)$ & \boldmath$\dcdx(t,\xdet)$ \\ \hline
    NECEEM & $\bar{c}\ \vr(x/l)$ & $0$ & $0$ \\ \hline
    cNECEEM & $0$ & $\bar{c}$ & $0$ \\ \hline
    SweepCE & $(\Leq,0,0)$ & $(0,\Teq,0)$ & $0$ \\ \hline
    sSweepCE & $(\Leq\vt(x-l),\ \Teq\vr(x/l),\ 0)$ & $0$ & $0$ \\ \hline
    sSweepCEEM & $(\Leq\vt(x-l),\ \Teq_1\vr(x/l)+\Teq_2\vt(x-l),\ \Ceq\vt(x-l))$ & $0$ & $0$ \\ \hline
    ECEEM & $(\Leq\vt(x-l),\ \Teq\vr(x/l),\ \Ceq\vt(x-l))$ & $(0,\Teq,0)$ & $0$ \\ \hline
    ppKCE & $(\Leq\vr((x-l_L)/l_T),\ \Teq\vr(x/l_T),\ 0)$ & $0$ & $0$ \\
    \hline
\end{tabular}
\end{table}
\egroup
\end{center}
\vspace{-1cm}
Note that $\xdet$ denotes the location of the experimental detector at the end of the capillary, and $\vt$ denotes the Heaviside function
\[ \vt(x) =\ \begin{cases}
    1\ \ \mathrm{if}\ x\geq 0 \\
    0\ \ \mathrm{otherwise}.
    \end{cases} \]

\newpage
\section{Solution of the Equations} \label{s03}

\subsection{Numerical Solution of the Complete Equations} \label{s0301}

\subsubsection{Discretization} \label{s030101}

As a reminder, the CCDR vector PDE equation is
\[ \dcdt + v\cdot\dcdx = D\cdot\ddcdx + R(c). \]
The standard way to discretize the equations would be
\[ \frac{c(t+\delt, x) - c(t,x)}{\delt} + v\cdot\frac{c(t,x+\delx)-c(t,x)}{\delx} = \]
\[ = D\cdot\frac{c(t,x-\delx) - 2 c(t,x) + c(t,x+\delx)}{\delx^2} + R(u) \]
yielding the vector iteration scheme
\[ c(t+\delt, x) = \frac{D\delt}{\delx^2}\cdot c(t,x-\delx) + \left(1+ \frac{v\delt}{\delx} - 2\frac{D\delt}{\delx^2}\right)\cdot c(t,x)\ + \]
\[ + \left(\frac{D\delt}{\delx^2} - \frac{v\delt}{\delx}\right)\cdot c(t,x+\delx) + R(c(t,x)) \delt. \]
This common scheme is unstable due to the fact that the coefficients can be negative for a sparse mesh, causing oscillations to emerge in this recursion, commonly referred to as instability \cite{versteeg2007introduction, roos2008robust, hundsdorfer2013numerical}. The main issue is that the P\'{e}clet number arising from the presence of both convection and diffusion (or the ratio $\max(v)/\min(D)$) is too large (convection-dominated case), requiring a very dense mesh for ensuring stability. This kind of instability can be partly resolved for convection--diffusion equations by weighting the diffusion term as in streamline upwind Petrov--Galerkin methods \cite{johnson2012numerical, galeao1988consistent, bank1990some, becker2007optimal}. Some programs have already been developed for modelling electrophoresis in other settings \cite{thormann2010dynamic, schwer1993computer, dutta2002numerical, mullerova2014twenty, staal2015versatile}.

Instead of a general partial resolution, we may attempt to resolve the instability by virtue of constant convection in our equations. Petrov et al. and others \cite{petrov2005kinetic, ermakov1992finite, fang2005general} suggest improving stability by aligning the grid points with the directions of plug peak motion in spacetime, by choosing special individual stepsizes $\delx_n = v_n\delt$. Upon further thought, we can eliminate the convection term entirely by first changing variables for each line of the equations as
\[ y_n\libeq x- v_n t,\ \ \vg_n(t,y_n)\libeq c_n(t,x) \]
resulting in the transformed equations
\[ \dt \vg_n = D_n\partial_{y_n}^2 \vg_n + R_n(\vg)\ \ (n = 1,\ldots, N). \]
Then the iterations will be executed separately for each concentration variable, and we interpolate between the meshes to compute the reaction term
\[ R_n(\vg) = R_n(\vg_1(t, y_1),\ldots,\ \vg_N(t, y_N)) = R_n(c(t,x)). \]
We discretize the $(t, y_n)$-planes as $\dely_n\libeq v_n\delt$ in order to align the $(t, y_n)$ grid points with the boundary conditions, which are now skewed in the transformed planes. So the iteration scheme becomes the following
\[ \vg_n(t+\delt, y_n) = \frac{D_n\delt}{\dely_n^2}\ \vg_n(t, y_n-\dely_n) + \left(1 - 2\frac{D_n\delt}{\dely_n^2}\right)\ \vg_n(t, y_n)\ + \]
 \[ +\ \frac{D_n\delt}{\dely_n^2}\ \vg_n(t, y_n+\dely_n) + R_n(\vg)\delt. \]

The stability of the iteration is now ensured, since the $v$ terms in the earlier discretization are eliminated, so the condition for stability becomes $\delt > \max_n(2D_n/v_n^2)$ which is easy to satisfy for a large P\'{e}clet number
\[ v_n\approx 10^{-3},\ D_n\approx 10^{-10}\ \rar\ \delt > 10^{-7}. \]
So with $t_{\mathrm{max}}=10^3\msec$, this implies a limit of at most $10^{10}$ grid points in time (and corresponding $\dely_n$). If $D_n=0$, then there is no limitation for refinement.

\subsubsection{The Solver Algorithm} \label{s030102}

The algorithm presented below finds approximate values of the concentration vector solution $c$ of the CCDR equations with $N\in\N$ lines, over a given grid $(t_i, x_j)\in\R_+^2\ (i= 0, 1,\ldots, I,\ j=0, 1,\ldots, J)$ where $x_J = \xdet$, with an equidistant temporal discretization $\delt$ and arbitrary values in space.

A different $x$-grid for each line of the equations is defined as $(t_i, x_{n,j}')$ where $x_{n,j}'\libeq j v_n \delt\ (j=0,1,\ldots,J_n)$ where $x_{n,J_n}' = \xdet$ (by choosing an equidistant spatial discretization $\delx_n= x_{n,j+1}'-x_{n,j}' = v_n\delt$). The $y$-grid for the transformed equations is defined as $(t_i, y_{n,i,j})$ where $y_{n,i,j}\libeq x_{n,j}' - v_n t_i\ (j=0,1,\ldots,J_n)$ (implying that $\dely_n = \delx_n$ for all $i$).

Denote the sought concentration values by $c_{n,i,j}\approx c_n(t_i, x_j)$, the values on the equidistant grid by $c_{n,i,j}'\approx c_n(t_i, x_{n,j}')$, while the values on the transformed grids by $\vg_{n,i,j}\approx \vg_n(t_i, y_{n,i,j})$. Note that $\vg_{n,i,j} = c_{n,i,j}'$, since $\vg_n(t,y_n)= c_n(t,x)$ by definition, for $y_n = x- v_n t$.

The reaction vector must be evaluated at the same spacetime grid points, so for each line of the equation $n\in\{1,\ldots, N\}$ we interpolate the concentration values on the other spatial grids $\vg_{k,i,j}\ (k\neq n,\ j=0,1,\ldots,J_k)$ onto the current $n$-th grid $(t_i, y_{n,i,j})\ (j=0,1,\ldots,J_n)$, and denote the new values as $\vg_{k,n,i,j}\ (j=0,1,\ldots,J_n)$ (note that we can define $\vg_{n,n,i,j}\libeq \vg_{n,i,j}$). The associated reaction values are denoted as
\[ R_{n,i,j}\libeq R_n(\vg_{1,n,i,j}, \ldots, \vg_{N,n,i,j})\ \ (j=0,1,\ldots,J_n). \]

The algorithm contains nested \textbf{for} loops, first according to time, then by each reactant. Therefore, it begins by evaluating the initial conditions from some initial condition function $c(0,x) = \mIC(x)$. The left boundary condition is given by the function $\vg_{n,i,0}\approx\vg_n(t_i, y_{n,i,0}) = c_n(t_i,0) = \mBC_n(t_i)$ and the right boundary condition is of Neumann type $\partial_{y_n}\vg_n(t_i, y_{n,i,J_n}) = 0$ implying that $\vg_{n,i,J_n} = \vg_{n,i,J_n-1}$. Note that we must also extrapolate the transformed concentration for $j=1$ as
\[ \vg_{n,i,j-1}\approx \frac{\vg_{n,i,j-2} + \vg_{n,i,j}}{2}\ \ \rar\ \ \vg_{n,i,j-2}\approx 2 \vg_{n,i,j-1} - \vg_{n,i,j}. \]
Lastly, denote the iteration coefficients (a convex combination) as
\[ A_n\libeq \frac{D_n\delt}{\dely_n^2},\ \ B_n\libeq 1 - 2 A_n\ \ (n= 1,\ldots,N). \]
\begin{algorithm}
\caption{(Numerical Solver for the CCDR Equations)} \label{s03010201}
\begin{algorithmic}[1]
\Function{NumSolCCDR}{$v, D, R, t, x$}
    \For{$n = 1,\ldots,N$}
        \For{$j = 0,1,\ldots,J_n$}
            \State $\vg_{n,0,j}\libeq \mIC_n(x_{n,j}')$
        \EndFor
        \State $c_{n,0,\cdot}\libeq$ \Call{Interpolate}{$\vg_{n,0,\cdot},\ x_{n,\cdot}',\ x$}
    \EndFor
    \For{$i = 0,1,\ldots,I-1$}
        \For{$n = 1,\ldots,N$}
            \For{$k = 1,\ldots,N,\ k\neq n$}
                \State $\vg_{k,n,i,\cdot}\libeq$ \Call{Interpolate}{$\vg_{k,i,\cdot},\ y_{k,i,\cdot},\ y_{n,i,\cdot}$}
            \EndFor
            \State $\vg_{n,i,0}\libeq \mBC_n(t_i)$
            \For{$j = 1,\ldots,J_n-1$}
                \State $R_{n,i,j}\libeq R_n(\vg_{1,n,i,j}, \ldots, \vg_{N,n,i,j})$
                \If{$j = 1$}
                    \State $\vg_{n,i+1,j}\libeq A_n (2 \vg_{n,i,j-1} - \vg_{n,i,j}) + B_n \vg_{n,i,j-1} + A_n \vg_{n,i,j} + R_{n,i,j} \delt$
                \Else
                    \State $\vg_{n,i+1,j}\libeq A_n \vg_{n,i,j-2} + B_n \vg_{n,i,j-1} + A_n \vg_{n,i,j} + R_{n,i,j} \delt$
                \EndIf
            \EndFor
            \State $\vg_{n,i+1,J_n}\libeq \vg_{n,i+1,J_n-1}$
            \State $c_{n,i+1,\cdot}\libeq$ \Call{Interpolate}{$\vg_{n,i+1,\cdot},\ x_{n,\cdot}',\ x$}
        \EndFor
    \EndFor
    \State \textbf{return}\ $c$
\EndFunction
\end{algorithmic}
\end{algorithm}

The algorithm may call various kinds of interpolation subroutines. Our implementation uses cubic Hermite splines, resulting in an $\mathcal{O}(1/I) = \mathcal{O}(\delt)$ convergence in the $L^2$ error from the exact solutions, as demonstrated computationally in Sec. \ref{s0403}. (Note that for the heat equation $\dcdt = D\ \ddcdx$ the error is well-known to be $\mathcal{O}(\delt) + \mathcal{O}(\delx^2)$ \cite{site008}.)

\subsection{The Exact Solution for a Simplification of NECEEM} \label{s0302}

We plan to explicitly solve a simplified case of the CCDR vector PDE equation
\[ \dcdt + v\cdot\dcdx = D\cdot\ddcdx + R(c) \]
with reaction function $R(c) = (\koff C,\ \koff C,\ -\koff C):\R_+^2\mto\R^3$ where $c = (L, T, C)$, via the method of fundamental solutions as demonstrated earlier for the case of a rectangular plug \cite{okhonin2004nonequilibrium}. Hereby the exact solutions are derived for a plug represented by a Gaussian initial condition, defined via the following density function
\[ \vr_G(x) = \vr_G[\mu, \vs^2](x)\libeq \frac{1}{\sigma\sqrt{2\pi}}\ \mathrm{exp}\left(-\frac{(x-\mu)^2}{2\sigma^2}\right). \]
The NECEEM vector initial condition according to Sec. \ref{s0202} is
\[ c(0,x) = \mIC(x)\libeq \bar{c}\ \vr_G(x/l) \]
while the boundary conditions remain unspecified for now, for the sake of simplicity.

Introducing some fundamental solutions will aid us in our derivation of an explicit solution. It can be shown by substitution that the equation
\[ \dt F + v\ \dx F - D\ \dx^2 F = -k F \]
with the initial condition $F(0,x) = \delta(x)$, is solved by
\[ F[k](t,x)\libeq \vt(t)\ \me^{-kt}\ \vr_G[v t,\ 2 D t](x). \]
On the other hand, the fundamental solution satisfying
\[ \dt F + v \dx F - D \dx^2 F = \delta \]
with a spatiotemporal Dirac delta function, is coincidentally the function $F[0]$.

Thus to find the third concentration component $C$ for some initial condition function $C(0,x)=\mIC_C(x)$, we must convolve it with the first fundamental solution above, resulting in
\[ C(t,x) = (\mIC_C\ast F_C[\koff](t,\cdot))(x) \]
where $F_C$ is defined similarly as above, with parameters $v_C$ and $D_C$.

In order to get the other two concentration components $L$ and $T$, we break them up into an equilibrium term (the solution of the $k=0,\ F(0,x)=\mIC(x)$ case above) and a dissipation term (the solution with right-hand side $\koff C$), resulting in
\[ L(t,x) = (\mIC_L\ast F_L[0](t,\cdot))(x) + \koff (C\ast F_L[0])(t,x) \]
\[ T(t,x) = (\mIC_T\ast F_T[0](t,\cdot))(x) + \koff (C\ast F_T[0])(t,x). \]
Since the above formulas include several function convolutions (single and double integrals), it is desirable to simplify our derivation with the following well-known identity
\[ \vr_G[\mu_1, \vs_1^2]\ast \vr_G[\mu_2, \vs_2^2] = \vr_G[\mu_1 + \mu_2,\ \vs_1^2 + \vs_2^2]. \]
Next, notice that for some $\mu_0, \vs_0 >0$ and $\mu\libeq \mu_0 l,\ \vs\libeq\vs_0 l$ we have
\[ \mIC(x) = \bar{c}\ \vr_G[\mu_0, \vs_0^2](x/l) = \bar{c} l\ \vr_G[\mu,\vs^2](x). \]
Employing the above general solution formulas for this special initial condition, we get
\[ C(t,x) = (\mIC_C\ast F_C[\koff](t,\cdot))(x) = l\Ceq \left(\vr_G[\mu,\vs^2]\ast \left(\vt(t)\me^{-\koff t} \vr_G[v_C t,\ 2 D_C t]\right)\right)(x) = \]
\[ = l \Ceq\ \vt(t)\ \me^{-\koff t}\ \vr_G[\mu + v_C t,\ \vs^2 + 2 D_C t](x). \]
With the above, we may now derive the other concentration components as follows
\[ L_{\meq}(t,x) = (\mIC_L\ast F_L[0](t,\cdot))(x) = l\Leq\ \vr_G[\mu+ v_L t,\ \vs^2 + 2 D_L t](x) \]
\[ L_{\mdis}(t,x) = \koff(C\ast F_L[0])(t,x) = \koff\int_{-\infty}^{+\infty} \left(C(\tau,\cdot)\ast F_L[0](t-\tau,\cdot)\right)(x)\ \md\tau = \]
\[ = \koff l \Ceq \int_{-\infty}^{+\infty} \vt(\tau) \me^{-\koff\tau} \left(\vr_G[\mu + v_C\tau, \vs^2 + 2 D_C\tau]\ast \vt(t-\tau)\vr_G[v_L(t-\tau), 2D_L(t-\tau)]\right)(x)\ \md\tau = \]
\[ = \koff l \Ceq \int_0^t \me^{-\koff\tau} \vr_G[\mu + v_L t + (v_C-v_L)\tau,\ \vs^2 + 2 D_L t + 2(D_C-D_L)\tau](x)\ \md\tau \]
\[ L = L_{\meq} + L_{\mdis} \]
and $T_{\meq},\ T_{\mdis},\ T$ are defined similarly.

Despite the extensive simplification of the exact solutions for the Gaussian case above -- otherwise containing even more computationally expensive convolution integrals -- we will show in Sec. \ref{s0402} that somewhat surprisingly, the numerical method of the previous section is much faster according to our computational tests. It is not only faster, but is capable of solving the CCDR equations for any reaction function and initial-boundary conditions, with a reasonable error according to Sec. \ref{s0403}.

\newpage
\section{Computational Results} \label{s04}

\subsection{The Direct Solver Package} \label{s0401}

The numerical solver algorithm of Sec. \ref{s030102} has been implemented in MATLAB for the KCE equations, though can be easily generalized to more reactants and any reaction mapping. The package is available on GitHub \cite{so002}. Currently the package contains only the direct solver for KCE, but the inverse solver built on it is also under development, serving our research described in the introduction.

The direct solver sub-package, contains an implementation of the numerical algorithm that is functional for all initial and boundary conditions, and plug types described in Sec. \ref{s0202}. Exact solutions have only been derived and implemented for the simplified NECEEM case in Sec. \ref{s0302} with Gaussian and rectangular plug types \cite{okhonin2004nonequilibrium}, and for MASKE \cite{okhonin2010maske} with any plug type.

Therefore, in order to compare the computational runtimes and errors between the numerical and exact solutions, we are restricted to either of the above two simplified methods. We chose simplified NECEEM with a Gaussian plug for the analysis below, since its equations are closer to the full KCE equations than those of MASKE, consisting of only two lines. Furthermore, the solution formulas can be minimized computationally as much as possible, by eliminating expensive convolution integrals which arise for other density functions. This way we achieve a lower bound on the computational complexity of the exact case, which can be compared with that of the numerical one. Specifically, three single convolution integrals are eliminated for $C, L_{\meq}, T_{\meq}$ and two double convolution integrals are reduced to single ones for $L_{\mdis}, T_{\mdis}$, as shown in Sec. \ref{s0302}.

The inverse problem of our research requires fast and accurate simulation of the solutions to the KCE equations for varying $\kon,\ \koff$ values, modeling experimental electropherogram signals \cite{vasskrylov2017kceinv}. Thus we analyze our direct solver in terms of both runtime and error below.

\subsection{Runtime Analysis} \label{s0402}

\begin{wrapfigure}{r}{0.625\textwidth}
\begin{center}
\includegraphics[width=0.6\textwidth]{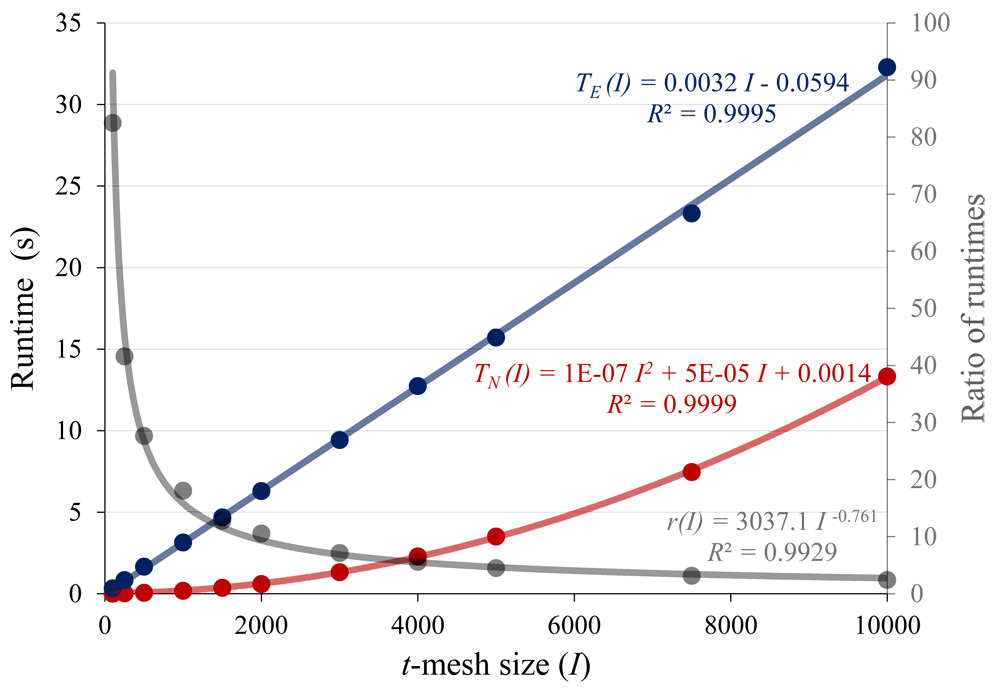}
\end{center}
\vspace{-0.3cm}
\caption{Average runtimes for the numerical solver (red), the exact solver (blue), and their ratio (grey).}
\label{s040201}
\end{wrapfigure}

In order to compare the runtime of our numerical solver (Sec. \ref{s030102}) to that of computing the exact solutions of simplified NECEEM (Sec. \ref{s0302}), we ran the solver for each temporal mesh size $I$ five times, and plotted the average of runtimes in Fig. \ref{s040201}. We are only interested in varying $I$ and not $J$ (the spatial mesh size), because the experimental signals are available only at the detector location $\xdet$, restricting the inverse problem to this single location in space. Test runs were done with an AMD A8-5550M processor, 8.00 GB of RAM, 64-bit Windows 10, and MATLAB R2016a, with the parameters
\[ \kon = 3500\mmcmol,\ \koff = 0.035,\ v = (3.3, 5, 4)\times 10^{-3}\mmsVsec,\ D = (7, 7, 7)\times 10^{-11}\mmssec \]
\[ \bar{c} = (1.86, 16.9, 3.14)\times 10^{-6}\mmolmc,\ l = 0.005\mmet,\ t_{\max} = 90\msec,\ \xdet = 0.2\mmet. \]

\newpage
The runtimes seem to closely follow quadratic and linear relationships for the numerical and exact solvers, respectively, with high coefficients of determination $R^2$. This is expected for the exact solver, which simply evaluates the formulas of Sec. \ref{s0302}. The quadratic polynomial for $T_N(I)$ makes sense, since the algorithm of Sec. \ref{s030102} must necessarily solve for all the nodes on the spacetime mesh to provide the values at the only relevant location $\xdet$. The transformed spatial meshes all depend on the temporal mesh linearly $\dely_n = v_n\delt$, implying that the spacetime mesh sizes must depend quadratically on $I$.

The exact solution formulas for simplified NECEEM (Sec. \ref{s0302}) provide a standard of comparison for the numerical solution values, since the formulas for $C, L_{\meq}, T_{\meq}$ are exact, while the formulas for $L_{\meq}, T_{\meq}$ can be approximated accurately with some quadrature method (we used the \textbf{integral} function of MATLAB, with a \textit{RelTol} value of $10^{-6}$).

It is important to observe that the ratio of runtimes $r(I)$ decreases for increasing $I$, and follows a power law with an exponent of $-0.761$ (though the ratio of a linear and a quadratic polynomial should asymptotically follow a hyperbola with an exponent of $-1$). This implies that the gain in runtime using the numerical solver is much larger for lower mesh sizes, the difference ranging between 1-2 orders of magnitude below $I = 2000$. For our end goal of an efficient inverse solver, a mesh size of $1000\leq I\leq 2000$ is sufficient.

Thus we have demonstrated that surprisingly the numerical solver has a significantly lower runtime than even an optimally low runtime of the exact solution formulas. For this to be relevant, the error between the solutions must be reasonable.

\subsection{Error Analysis} \label{s0403}

\begin{figure}[H]
\centering
\includegraphics[width=300pt]{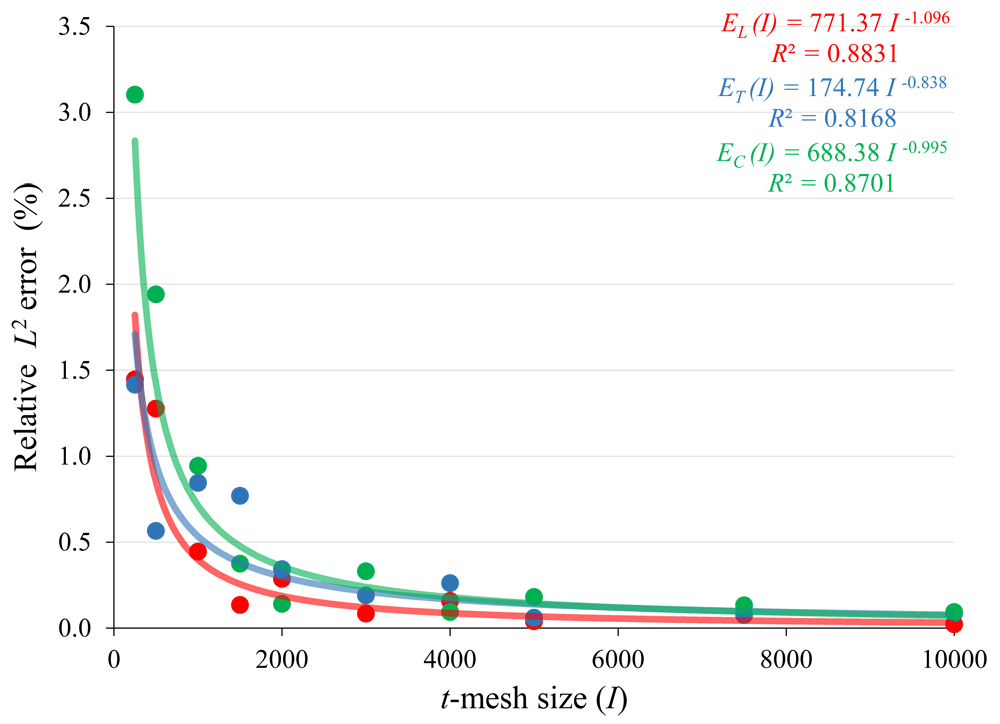}
\caption{Plot of the relative $L^2$ error between the exact and numerical solution signals at the detector, for increasing temporal mesh sizes.}
\label{s040301}
\end{figure}

To show that the numerical solutions generated by Algorithm \ref{s03010201} are ``reasonable'', we demonstrate that the $L^2$-error from the exact values vanishes with an increasing temporal mesh size $I$. Fig. \ref{s040301} shows that the error decreases for increasing temporal mesh size according to similar power laws for the three concentration components. Taking the weighted average of the three exponents according to the coefficients of determination ($R^2$ values) gives $-0.9798$. Thus we conjecture that all three error components are inversely proportional to the temporal mesh size for large $I$, i.e. precisely of order $\mathcal{O}(1/I) = \mathcal{O}(\delt)$ (Sec. \ref{s030102}).

\begin{figure}[H]
\centering
\vspace{0.5cm}
\includegraphics[width=440pt]{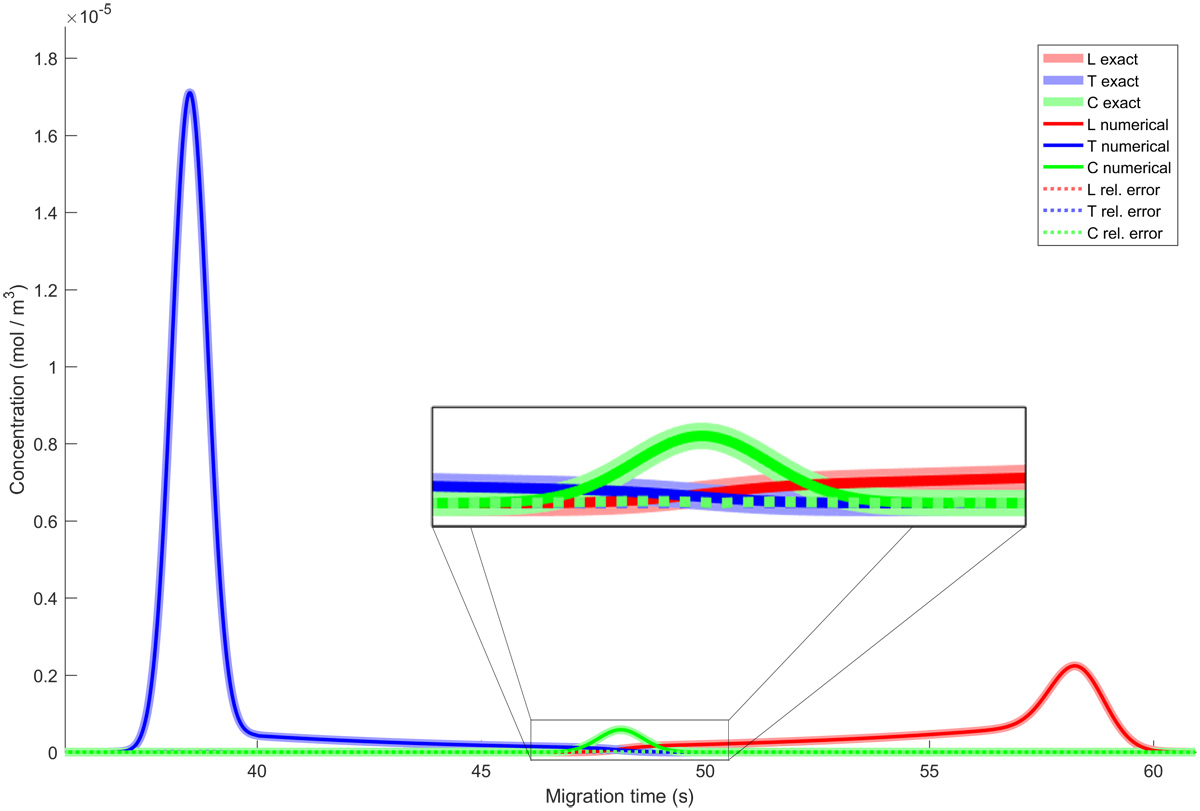}
\caption{The exact and numerical solutions for the parameters in Sec. \ref{s0402}, and temporal mesh size $I = 10^4$.}
\label{s040302}
\end{figure}

%\begin{mdframed}[style=mystyle]
The convergence of Algorithm \ref{s03010201} for increasing mesh size is self-evident from its design. Clearly, the error in the solution -- arising from the derivative approximations, and the multimesh interpolation -- vanishes as the spacetime mesh size increases. For an infinitesimally fine mesh, the error is necessarily zero. Thus the above comparison on Fig. \ref{s040301} -- between the exact and simulated solutions in the simplified case -- is not intended to be a proof of convergence, but merely a rate analysis of the de facto vanishing of the error with respect to mesh size. It gives an idea of how ``reasonable'' this convergence is.

Nevertheless, this computational error analysis of the simplified system bears relevance to the original equations. The two differ in the reaction terms: (1) original: $-\kon LT + \koff C$; (2) simplified: $\koff C$, but the solutions differ negligibly. As illustrated on Fig. \ref{s040302} for the simplified solution, the peaks are separated in a way that the following hold
\[ LT \gg 0\ \ \mathrm{when}\ \ C \gg 0;\ \ \ LT \approx 0\ \ \mathrm{when}\ \ C \approx 0;\ \ \ LT \ll C\ \ \mathrm{when}\ \ C \gg 0. \]
Consequently, the following approximations hold
\[ -\kon LT + \koff C \approx 0\ \ \mathrm{when}\ \ C \approx 0;\ \ \ -\kon LT + \koff C \approx \koff C\ \ \mathrm{when}\ \ C \gg 0. \]
So we may conclude heuristically, that the above rate analysis of the error in the simplified case, is closely aligned with the original.
%\end{mdframed}

%\newpage
\section{Concluding Remarks} \label{s05}

\subsection{Summary and Future Directions} \label{s0501}

A stable algorithm has been introduced for the efficient generation of accurate solutions to a set of convection--diffusion equations, with constant convection velocities. Stability has been demonstrated algebraically, due to the change of variables to new spatial meshes, for each line of this system of partial differential equations. Both efficiency and accuracy have been demonstrated as well.

The direct solver package corresponding to this work \cite{so002} may be extended in two potential directions: the implementation of new initial and boundary conditions, perhaps with a generic classification system; or the derivation of new exact solutions for either other simplifications of the original equations, or perhaps the explicit solution of the complete equations. According to our runtime analysis, however, the computation of closed form solutions is not likely to be more efficient, even in an extremely simple Gaussian case which gives a lower bound on computational complexity, implying the permanent significance of our solver. % the introduced numerical solver.

The relevance of this algorithm to our further work on an experimental inverse problem has also been highlighted. Indeed, this is the direction our solver package is being extended in, now enabled by the fast computation of parametrized solutions with an error measured from an experimental signal, which is minimized via an optimization algorithm.

\subsection{Applicability to Other Physical Models} \label{s0504}

The Kinetic Size-Exclusion Chromatography (KSEC) model \cite{bao2014kinetic, bao2015pre} can be written under certain experimental conditions in a form quite similar to the KCE equations \cite{cherney2016slow, cherney2011slow}. However, our program package in its current form \cite{so002} cannot handle the KSEC equations, mainly due to additional factors in the reaction term and the initial conditions, even in a simplified case \cite{cherney2016slow}. Furthermore, due to their more special form, the KSEC equations can be reduced to an equivalent form, for which a more efficient solver should be developed.

Algorithm \ref{s03010201} can also be adopted to the ACE model \cite{kanoatov2014extracting} and a separation-based approach \cite{petrov2011separation}, though it is no doubt applicable to the simulation of many other physical processes.

\subsection{Experimental Remarks} \label{s0503}

Both the KCE \cite{krylov2007kinetic} and KSEC \cite{bao2014kinetic} equations describe kinetic separation methods, with concentration components as functions of time and space, where a reversible binding reaction is paralleled by the separation of reactants, convected at constant velocities in a capillary.

%\begin{mdframed}[style=mystyle]
%Lastly, we wish to make some remarks that are potentially relevant to experimenters.
The presented discretization method was developed primarily for the KCE equations, which describe the migration of and interaction between three species: the ligand $L$, the target $T$, and the complex $C$. The KCE equations do not incorporate a variety of other ``background'' species and reactions -- such as components of the acid--base equilibration in the background electrolyte -- which occur during electrophoresis of these three species.

In practice, the KCE method utilizes experimentally-measured velocities of the three species, and plug lengths corresponding to them. Such parameters are sufficient for the description of the relevant migration--reaction phenomena, only if the electrophoretic experiment is planned and carried out properly.

Most importantly, the concentrations of the analytes in KCE should be well-below the concentrations of the ions in the background electrolyte. In turn, in KSEC experiments, the equilibration between the pores and the free volume should be fast. Beneficially, these conditions are easily satisfied for both KCE and KSEC. Analyte concentrations are typically at least four orders of magnitude below the background electrolyte concentration, while the sub-nanometer size of pores allows fast equilibration between them and the free volume.
%Beneficially, this condition is easily satisfied for KCE, as analyte concentrations are typically at least four orders of magnitude below the background electrolyte concentration.
%\end{mdframed}

%Both the KCE and KSEC equations describe kinetic separation methods, with concentration components as functions of time and space, where a reversible binding reaction is paralleled by the separation of reactants, convected at constant velocities in a capillary. %In KSEC experiments, the equilibration between the pores and the free volume should be fast.

%Beneficially, the above conditions can be easily satisfied for both KCE and KSEC, as analyte concentrations are typically at least four orders of magnitude below the background electrolyte concentration and the sub-nanometer size of pores allows fast equilibration between the pores and the free volume.

\subsection{Acknowledgements} \label{s0502}

This work was supported by the Natural Sciences and Engineering Research Council of Canada (grant: CRDPJ 485321-15). We are grateful to Mirzo Kanoatov for noticing implementation bugs in the direct solver package, and for his suggestions towards improving its user-friendliness. We also appreciate the valuable suggestions made by the referees.

\newpage
\bibliographystyle{abbrv}
\bibliography{mybib2}
\addcontentsline{toc}{section}{\textbf{References}}

\end{document}